\documentclass[aps,pra,amssymb,amsmath,eqsecnum,nofootinbib]{revtex4}
\usepackage{graphicx}

\newtheorem{definition}{Definition}[section]
\newtheorem{theorem}{Theorem}[section]
\newtheorem{proposition}{Proposition}[section]

\newtheorem{corollary}{Corollary}[section]

\newtheorem{remark}{Remark}[section]

\newenvironment{hypothesis}{HP: \begin{center}} {\end{center}}
\newenvironment{thesis}{TH: \begin{center}} {\end{center}}
\newenvironment{proof}{\begin{center}PROOF: \end{center}} {$ \blacksquare $}

\begin{document}
\title{There exist consistent temporal logics admitting changes of History}
\author{Gavriel Segre}
\begin{abstract}
Introducing his Chronology Protection Conjecture Stephen Hawking
said that it seems that there exists a Chronology Protection
Agency making the Universe safe for historians.

Without taking sides about such a conjecture we show that the
Chronology Protection Agency is not necessary in order to make the
Universe unsafe for historians but safe for logicians.
\end{abstract}
\maketitle
\newpage
\tableofcontents
\newpage
\section{Introduction}
 Given a time-orientable space-time $ ( M , g_{ab}
) $ let us recall that \cite{Hawking-Ellis-73}, \cite{Wald-84}:
\begin{definition} \label{def:chronology violating set}
\end{definition}
\emph{chronology violating set of $ ( M , g_{ab} )$:}
\begin{equation}
    V_{chronology} ( M , g_{ab} ) \; := \; \cup_{p \in  M} I^{+} (p) \cap I^{-} (p)
\end{equation}
\begin{definition} \label{def:chronology horizon}
\end{definition}
\emph{chronology horizon of $ ( M , g_{ab} )$:}
\begin{equation}
   H_{chronology} ( M , g_{ab} ) \; := \; \partial [ I^{+} (  V_{chronology} ( M , g_{ab} )
   )]
\end{equation}

 A curious feature of General Relativity is that there
exist solutions $( M , g_{ab} ) $ of Einstein's equation such that
$ V_{chronology} ( M , g_{ab} ) \; \neq \; \emptyset $.

The so called time travel paradoxes then occur.

These paradoxes can be divided in two classes:
\begin{itemize}
    \item \emph{consistency paradoxes} involving the effects of
    the changes of the past (epitomized by the celebrated
    Grandfather Paradox in which a time-traveller goes back in the
    past and prevents the meeting of his grandfather and his
    grandmother)
    \item \emph{bootstrap paradoxes} involving the presence of
    loops in which the source of the production of some
    information disappears (as an example let us suppose that
    Einstein learnt  Relativity Theory from
    \cite{Hawking-Ellis-73},
    \cite{Wald-84} given to him by a time-traveller gone back to
    1904).
\end{itemize}

They has been faced by the scientific community in different ways
(see the fourth part "Time Travel" of \cite{Visser-96} as well as
\cite{Visser-03}):
\begin{enumerate}
    \item adding to General Relativity some ad hoc axiom
    precluding the physical possibility of causal loops (such as
    the strong form of Penrose's Cosmic Censorship Conjecture)
    \item appealing to consistency conditions (such as in Novikov's
    Consistency Conjecture) requiring that causal loops, though allowing causal influence on the past, don't allow
    alteration of the past
    \item arguing that the problem is removed at a quantum level (such as in Hawking's Chronology
    Protection Conjecture \cite{Hawking-92} stating that the  classical possibilities to implement time-travels are  destroyed by quantum effects)
    \item arguing that the so called time-travel paradoxes are only apparent and
    may be bypassed in a mathematical consistent way
\end{enumerate}

As to General Relativity we think that discarding tout court non
globally-hyperbolic solutions of Einstein's equation considering
them "unphysical" is a conceptually dangerous operation:

in presence of "unphysical" solutions of physical equations we
have always to remember that our intuition is not a neutral
quality but is affected by the Physics to which we are used.

Consequentially, in presence of new Physics, it is natural that it
 appears to us as counter-intuitive.

If Dirac had discarded as "unphysical"  the negative-energy
 solutions of his equation  he would have never predicted the
 existence of anti-matter \cite{Weinberg-95}.

 \smallskip

 While we agree with Visser's \cite{Visser-03} idea that
 solutions with "non-localized" chronology violating set may be seen
 as produced by a sort of\emph{ garbage in-garbage out} phenomenon (where
 the \emph{garbage in} are perverse initial-value conditions and the
 \emph{garbage-out} are the resulting perverse solutions) we think that
 solutions with "localized" chronology violating set should be taken
 seriously.

 Of course, depending on the precise mathematical definition that we
 adopt for the term "localized", we may arrive to different
 conclusions.

 A minimal definition of the term "localized" would consist in
 imposing that $  V_{chronology} ( M , g_{ab} ) \neq M $.

 This is sufficient to discard G\"{o}del's solution, the Van
 Stockum - Tipler time machine, some spinning cosmic string
 time machine but not Gott time machine that may be excluded only
 assuming a more restrictive definition of "localized" as
 "suitably bounded".

\smallskip

An other  argument often used in the literature consists in the
refutal as "unphysical" of any non asymptotically-flat space-time;
this is (at least) curious: the fact that asymptotically flatness
is a condition required in order to be able to define a black-hole
(as the complement of the causal past of future null infinity $
\mathcal{B} \, := \, M - J^{-} ( \mathcal{I}^{+} ) $) is not a
good reason to assume as "physicality"'s criterion one
incompatible with the assumption of homogeneity and isotropy under
which   the Friedmann-Robertson-Walker solutions of Classical
Cosmology are derived.

\smallskip

Finally we  agree with the Headrick-Gott's \cite{Headrick-Gott-94}
refutation of  the claim that the observed absence of
time-travellers in the present and in the past would be an
empirical datum supporting the physical impossibility of
time-travel:

no experimental fact on $ M - H_{chronology} ( M , g_{ab} ) $ can
be used as an argument in favour or against the hypothesis that $
V_{chronology} ( M , g_{ab} ) \neq \emptyset $.

\bigskip

When also Quantum Mechanics is taken into account we again agree
with Visser's viewpoint \cite{Visser-03} according to which the
Kay-Radzikowski-Wald singularity theorems (stating that in
presence of a non-empty chronology-violating set there are points
of the \emph{chronology horizon} where  the two-point function is
not of Hadamard form \cite{Kay-Radzikowski-Wald-97},
\cite{Kay-06}) has to be interpreted not as a support of Hawking's
Chronology Protection Conjecture but as an evidence of the fact
that in such a situation Semi-classical Quantum Gravity (defined
as Quantum Field Theory on a fixed curved background augmented
with the semi-classical Einstein equation $ R_{a b} - \frac{1}{2}
R g_{a b} \; = 8 \pi   < \psi | \hat{T}_{ab} | \psi > $ taking
into account the back-reaction of the quantum fields on the
spacetime's geometry \cite{Wald-94}) is not a good approximation
of Quantum Gravity.

Hence we think that the status of Hawking's Chronology Protection
Conjecture may be decided only at the Quantum Gravity level.

 Since most of
the phenomenology of Quantum Gravity is detectable only at the
Planck scale (lengths of the magnitude of the Planck length $
l_{Planck} \sim 10^{- 33}
 \, cm$, energies of the magnitude of the Planck energy $E_{Planck} \sim 10^{19} \, GeV $) that is enormously far from our present possibility of experimental
investigation, the quantum status of Hawking's
Chronology-Protection Conjecture may be at present investigated
only on a theoretical basis, all the rival alternative proposals
(String Theory
\cite{Deligne-Etingof-Freed-Jeffrey-Kazhdan-Morgan-Morrison-Witten-99a},
\cite{Deligne-Etingof-Freed-Jeffrey-Kazhdan-Morgan-Morrison-Witten-99b},
Loop Quantum Gravity \cite{Rovelli-04}, Connes' Quantum Gravity
\cite{Connes-98}, \cite{Landi-97},  Simplicial Quantum Gravity
\cite{Ambiorn-Carfora-Marzuoli-97}, Prugovecki's Quantum Gravity
\cite{Prugovecki-92}, Finkelstein's Quantum Gravity
\cite{Finkelstein-97}, $ \cdots $) being far from giving, on this
issue, clear and univocal answers.

\smallskip

Without taking sides about the physical possibility of chronology
violations, in this paper we will show that, from a logical
viewpoint, the so-called time travel paradoxes may be seen as a
consequence of the fact that, in presence of chronology
violations, we insist on the adoption of the usual Temporal Logic.

The adoption of suitable unusual Temporal Logics allows to get rid
of any mathematical inconsistency.

Indeed not only the affection of the past but even its change can
be formalized by  suitable consistent Temporal Logics.
\newpage
\section{Temporal Logics}

Temporal Logics are the particular Modal Logic in which the
following temporal operators are introduced \cite{Prior-62}:
\begin{itemize}
    \item P := "it has been true that"
    \item H := "it has always been true that"
    \item F := "it will be true that"
    \item G := "it will always be true that"
\end{itemize}

The absence of any temporary operator in a formula in Temporary
Logic means that it is temporally referred to the present time:
\begin{equation}
    x \; := \; \text{"x is true at the present time"}
\end{equation}

To avoid too many parenthesis we will assume that the temporal
operators P,H,F,G, are at the top of the strength's hierarchy of
Propositional Logic introduced in the appendix \ref{sec:Classical
Propositional Logic}.

\begin{remark}
\end{remark}
We adhere to Arthur Prior's traditional notation.

From a conceptual viewpoint is anyway important to remark that:
\begin{enumerate}
    \item P is the past existential quantificator. This fact may
    be remarked introducing the notation:
\begin{equation}
    \exists_{-} \; := P
\end{equation}
    \item H is the past universal quantificator. This fact may
    be remarked introducing the notation:
\begin{equation}
    \forall_{-} \; := H
\end{equation}
    \item F is the future existential quantificator. This fact may
    be remarked introducing the notation:
\begin{equation}
    \exists_{+} \; := F
\end{equation}
    \item G is the future universal quantificator. This fact may
    be remarked introducing the notation:
\begin{equation}
    \forall_{+} \; := G
\end{equation}
\end{enumerate}

\smallskip

In the case in which one assume a nonrelativistic notion of
absolute time taking values on $ \mathbb{R} $ and identifying
conventionally the present time with $ t = 0$ the temporal
operators may be defined considering time-dependent propositions
of the form:
\begin{equation}
    x(t) \; := \; \text{"The proposition x is true at time t"}
\end{equation}
and defining them as:
\begin{definition} \label{def:temporal operators}
\end{definition}
\begin{equation}
    P x \; := \; \exists t \in ( - \infty , 0 ) \, : \, x(t)
\end{equation}
\begin{equation}
    H x \; := \;  x(t) \; \forall t \in ( - \infty , 0 )
\end{equation}
\begin{equation}
    F x \; := \; \exists t \in ( 0 , + \infty  ) \, : \, x(t)
\end{equation}
\begin{equation}
    G x \; := \;  x(t) \; \forall t \in ( 0 , +  \infty  )
\end{equation}
\begin{equation}
    x \; := \; x(0)
\end{equation}

\smallskip

\begin{remark} \label{rem:conditions in which the temporalized definitions of the temporal operators can be adopted}
\end{remark}
Assuming a nonrelativistic notion of time with a nontrivial
topological structure (such as, for instance, those discussed in
\cite{Segre-06}) or a relativistic notion of time one has to give
up the definition \ref{def:temporal operators}

\smallskip

\begin{remark}
\end{remark}
Given a formula containing a composed temporal operator of the
form $ x y \; \; x,y \in \{P,H,F,G \} $ let us observe that x
alterates the present time for y.

In the case in which the the definition \ref{def:temporal
operators} may be adopted this fact can be expressed by the
following:

\begin{proposition} \label{prop:time-dependent expression of the composed temporal operators}
\end{proposition}

\begin{hypothesis}
\end{hypothesis}
\begin{center}
 The definition \ref{def:temporal operators} is assumed
\end{center}

\begin{thesis}
\end{thesis}

\begin{enumerate}
    \item
\begin{equation}
    P P x \; = \; \exists t_{1} \in ( - \infty , 0 ) : ( \exists
    t \in ( - \infty , t_{1} ) : x(t) )
\end{equation}
   \item
\begin{equation}
    P H x \; = \; \exists t_{1} \in ( - \infty , 0 ) : ( x(t) \;
    \forall t \in ( - \infty , t_{1} ) )
\end{equation}
 \item
\begin{equation}
    P F x \; = \; \exists t_{1} \in ( - \infty , 0 ) : ( \exists
    t \in ( t_{1} , + \infty ) : x(t))
\end{equation}
 \item
\begin{equation}
    P G x \; = \; \exists t_{1} \in ( - \infty , 0 ) : ( x(t) \;
    \forall t \in ( t_{1} , + \infty ))
\end{equation}
 \item
\begin{equation}
    H P x \; = \;  ( \exists t \in  ( - \infty , t_{1} ) : x(t) )     \; \forall t_{1} \in ( - \infty , 0 )
\end{equation}
    \item
\begin{equation}
    H H x \; = \; ( x(t) \; \forall t \in ( - \infty , t_{1} ) )
    \; \forall t_{1} \in ( - \infty , 0 )
\end{equation}
  \item
\begin{equation}
    H F x \; = \;  ( \exists t \in ( t_{1} , + \infty ) : x(t) )   \; \forall t_{1} \in ( - \infty , 0 )
\end{equation}
  \item
\begin{equation}
    H G x \; = \;  (  x(t) \; \forall t \in ( t_{1} , + \infty ) ) \; \forall t_{1} \in ( - \infty , 0 )
\end{equation}
 \item
\begin{equation}
    F P x \; = \; \exists t_{1} \in ( 0 , + \infty ) : ( \exists t \in ( - \infty , t_{1} ) :  x(t)   )
\end{equation}
    \item
\begin{equation}
    F H x \; = \; \exists t_{1} \in ( 0 , + \infty ) :  ( x(t) \;
    \forall t \in ( - \infty , t_{1} ))
\end{equation}
  \item
\begin{equation}
    F F x \; = \; \exists t_{1} \in ( 0 , + \infty ) : (   \exists t \in ( t_{1} , + \infty
    ) : x(t) )
\end{equation}
   \item
 \begin{equation}
    F G x \; = \;  \exists t_{1} \in ( 0 , + \infty ) :  ( x(t) \;
    \forall t \in ( t_{1} , + \infty ))
\end{equation}
  \item
 \begin{equation}
    G P x \; = \;   (  \exists t \in ( - \infty , t_{1} ) : x(t)  ) \; \forall t_{1} \in ( 0 , + \infty )
\end{equation}
 \item
\begin{equation}
    G H x \; = \;  ( x(t) \; \forall t \in ( - \infty , t_{1} ) )  \; \forall t_{1} \in ( 0 , + \infty )
\end{equation}
 \item
 \begin{equation}
    G F x \; = \;  \;  ( \exists t \in ( t_{1} , + \infty ) :  x(t) )  \forall t_{1} \in ( 0 , + \infty )
\end{equation}
    \item
\begin{equation}
    G G x \; = \; ( x(t) \; \forall t \in ( t_{1} , + \infty )) \;
    \forall t_{1} \in ( 0 , + \infty )
\end{equation}
\end{enumerate}
\begin{proof}
 The thesis follows by multiple application of the definition \ref{def:temporal
 operators}
\end{proof}

\smallskip

Let us observe that the temporal operators H and G may be defined
in the following way:
\begin{definition}
\end{definition}
\begin{equation}
    H x \; := \; \neg P \neg x
\end{equation}
\begin{equation}
    G x \; := \; \neg F \neg x
\end{equation}

\bigskip

Let us now introduce the following basic:

\begin{definition} \label{def:Temporal Logic}
\end{definition}
\emph{Temporal Logic:}
\begin{center}
  a modal logic
obtained adding to Classical Propositional Logic (see appendix
\ref{sec:Classical Propositional Logic}) the temporal operators P,
H, F, G  and suitable axioms and inference rules governing their
behavior among which there are the following:
\begin{equation} \label{eq:axiom for the past}
AXIOM_{-}  \; :=  \;  H x \, \rightarrow \, P x
\end{equation}
\begin{equation} \label{eq:axiom for the future}
AXIOM_{+}  \; :=  \;  G x \, \rightarrow \, F x
\end{equation}
\end{center}

\smallskip

\begin{remark}
\end{remark}
Given a temporal logic it is important to keep attention about the
meaning given to a logical variable x appearing in a formula.

We will assume that x will continue to have the same meaning it
had in Propositional Calculus (defined by the definition
\ref{def:logical notation}).

Consequentially x will not be allowed to contain temporal
operators.

\smallskip

\begin{remark}
\end{remark}
Let us remark that if the definition \ref{def:temporal operators}
is assumed the axiom  $ AXIOM_{-} $ and $ AXIOM_{+} $ become
trivially provable statements.

\section{Changing the past}

Given a temporal logic T:

\begin{definition} \label{def:temporal logic admitting changes of History}
\end{definition}
\emph{T admits changes of History:}
\begin{enumerate}
    \item "The fact that x is now true doesn't imply that it
    will be always true that x has been true"
\begin{equation}
     \neg (  x \; \rightarrow \; G P x)
\end{equation}
    \item "The fact that x has been true doesn't imply that it
    will be always true that x has been true"
\begin{equation}
    \neg ( P x \; \rightarrow \; G P x)
\end{equation}
 \item "The fact that x has always been true doesn't imply that it
    will be always true that x has been true"
\begin{equation}
    \neg ( H x \; \rightarrow \; G P x)
\end{equation}
  \item "The fact that x is now true doesn't imply that it
    will be true that x has been true"
\begin{equation}
    \neg ( x \; \rightarrow \; F P x)
\end{equation}
 \item "The fact that x has been true doesn't imply that it
    will be true that x has been true"
\begin{equation}
    \neg ( P x \; \rightarrow \; F P x)
\end{equation}
  \item "The fact that x has always been true doesn't imply that
  it will be true that x has  been true"
\begin{equation}
    \neg ( H x \; \rightarrow \; F P x)
\end{equation}
\end{enumerate}

Let us remark that:
\begin{proposition}
\end{proposition}

\begin{hypothesis}
\end{hypothesis}
\begin{enumerate}
    \item T temporal logic
    \item the definition \ref{def:temporal operators} is assumed
\end{enumerate}
\begin{thesis}
\end{thesis}
\begin{center}
 T doesn't admit changes of History
\end{center}
\begin{proof}
\begin{enumerate}
    \item
\begin{equation}
   x \; = \; x(0)
\end{equation}
while:
\begin{equation}
  GP x \; = \;  (  \exists t \in ( - \infty , t_{1} ) : x(t)  ) \; \forall t_{1} \in ( 0 , + \infty
  )
\end{equation}
that considering in the right hand side the value $ t=0 $ implies
that:
\begin{equation}
    x \; \rightarrow \; G P x
\end{equation}
that applying the $ AXIOM_{+} $ of definition \ref{def:Temporal
Logic} and the theorem \ref{th:transitivity of implication}
implies that:
\begin{equation}
  x \; \rightarrow \; F P x
\end{equation}
    \item
\begin{equation} \label{eq:auxiliary 1}
    P x \; = \;  \exists t \in ( - \infty , 0 ) \, : \, x(t)
\end{equation}
while:
\begin{equation} \label{eq:auxiliary 2}
  GP x \; = \;  (  \exists t \in ( - \infty , t_{1} ) : x(t)   ) \; \forall t_{1} \in ( 0 , + \infty
  )
\end{equation}
that considering as particular value of t in the right-hand side
of  eq. \ref{eq:auxiliary 2} the one whose existence is stated by
the right hand side of eq. \ref{eq:auxiliary 1} implies that:
\begin{equation}
    P x \; \rightarrow \; G P x
\end{equation}
that applying the $ AXIOM_{+} $ of definition \ref{def:Temporal
Logic} and the theorem \ref{th:transitivity of implication}
implies that:
\begin{equation}
  Px \; \rightarrow \; F P x
\end{equation}
    \item
\begin{equation}
    H x \; = \;  x(t) \; \forall t \in ( - \infty , 0 )
\end{equation}
so that considering any value  $ t \in ( - \infty , 0 ) $ as the t
in the right hand side of eq. \ref{eq:auxiliary 2} implies that:
\begin{equation}
    H x \; \rightarrow \; G P x
\end{equation}
that applying the $ AXIOM_{+} $ of definition \ref{def:Temporal
Logic} and the theorem \ref{th:transitivity of implication}
implies that:
\begin{equation}
    H x \; \rightarrow \; F P x
\end{equation}
\end{enumerate}
\end{proof}

\bigskip

Let us now prove two propositions that will be our key ingredients
in analyzing the consistency of temporal histories admitting
changes of History.

\begin{proposition} \label{prop:on the temporalized Principle of Contradiction}
\end{proposition}

\begin{hypothesis}
\end{hypothesis}
\begin{enumerate}
    \item T temporal logic
    \item T admits changes of History
\end{enumerate}

\begin{thesis}
\end{thesis}
\begin{center}
"The fact that the Principle of Contradiction now holds doesn't
imply that it will always be true that the Principle of
Contradiction held ":
\end{center}
\begin{equation}
   \neg [ \neg ( x \wedge \neg x ) \; \rightarrow \; GP  \neg ( x \wedge \neg x
   )]
\end{equation}
\begin{proof}
  The thesis immediately follows applying the definition \ref{def:temporal logic admitting changes of
  History} to the Principle of Noncontradiction (the theorem \ref{th:Principle of noncontradiction}) holding at the
  present time
\end{proof}

\smallskip

The conceptual reason why the Principle of Noncontradiction is so
important in Propositional Logic (as well as in Predicative Logic
of $ 1^{th}$ and $ 2^{th} $ order) derives by the Scotus Theorem
(the theorem \ref{th:Scotus theorem}) asserting that \emph{"ex
absurdo quodlibet sequitur"}, i.e. that every proposition can be
inferred starting from a contradiction.

The situation is deeper in a temporal logic admitting changes of
History:
\begin{proposition} \label{prop:on the temporalized Scotus theorem}
\end{proposition}
\begin{center}
 "The fact that everything may be proved from a contradiction doesn't imply that
 it will always be true that everything could be proved from a
 contradiction"
\end{center}

\begin{hypothesis}
\end{hypothesis}
\begin{enumerate}
    \item T temporal logic
    \item T admits changes of History
\end{enumerate}
\begin{thesis}
\end{thesis}
\begin{equation}
   \neg [ ( x_{1} \wedge \neg x_{1} \rightarrow x_{2} ) \; \rightarrow \; GP  ( x_{1} \wedge \neg x_{1} \rightarrow x_{2} )]
\end{equation}
\begin{proof}
  The thesis immediately follows applying the definition \ref{def:temporal logic admitting changes of
  History} to the Scotus Theorem (the theorem \ref{th:Scotus theorem}) holding at the
  present time
\end{proof}

\smallskip

Let us now analyze the notion of consistency of a temporal logic.

Given a  temporal logic T:

\begin{definition} \label{def:consistency}
\end{definition}
\emph{T is consistent:}
\begin{equation}
  \neg ( x \wedge \neg x)
\end{equation}

Then:
\begin{proposition}
\end{proposition}
\begin{center}
 every temporal logic is consistent
\end{center}
\begin{proof}
The thesis is nothing but theorem \ref{th:Principle of
noncontradiction} of Propositional Logic.
\end{proof}
\smallskip

\begin{definition} \label{def:always consistency}
\end{definition}
\emph{T is always-consistent:}
\begin{equation}
 [\neg ( x \wedge \neg x) ] \wedge [ H \neg ( x \wedge \neg x) ] \wedge [ G \neg ( x \wedge \neg x)]
\end{equation}

\smallskip

Proposition\ref{prop:on the temporalized Scotus theorem} shows
that always-consistency is not necessary to exorcize the \emph{"ex
absurdo quodlibet sequitur"} phenomenon.

Anyway even imposing always-consistency changes of History are not
banned:
\begin{proposition}
\end{proposition}
\begin{equation}
    \exists \; T \text{always-consistent temporal logic admitting changes of History}
\end{equation}
\begin{proof}
It is sufficient to take the axioms of T such that:
\begin{center}
"the fact that the fact that Principle of Contradiction now holds
doesn't imply that it will always be true that the Principle of
Contradiction held doesn't imply that the Principle of
Contradiction didn't hold":
\end{center}
\begin{equation}
   \neg \{  \neg [ \neg (x
   \wedge \neg x ) \rightarrow GP \neg ( x
   \wedge \neg x ) ]  \; \rightarrow \;  P (  x
   \wedge \neg x) \}
\end{equation}
\end{proof}

\newpage

\appendix
\section{Classical Propositional Logic} \label{sec:Classical Propositional Logic}

In this section we will briefly review the basic notions of
Propositional Calculus in its boolean formulation
\cite{Mendelson-64}:

 Given the binary
alphabet $ \Sigma := \{ 0 , 1 \} $ (where 1 is identified with
"true" and 0 is identified with "false") let us introduce the
following logical operators:

\begin{definition}
\end{definition}
\emph{negation:}

$ \neg : \Sigma \mapsto \Sigma $:

\begin{equation}
    \neg 0 \; := \; 1
\end{equation}
\begin{equation}
    \neg 1 \; := \; 0
\end{equation}

\smallskip

\begin{definition}
\end{definition}
\emph{conjunction:}

$ \wedge : \Sigma \times \Sigma \mapsto \Sigma$:
\begin{equation}
    0 \wedge 0 \; := \; 0
\end{equation}
\begin{equation}
    0 \wedge 1 \; := \; 0
\end{equation}
\begin{equation}
    1 \wedge 0 \; := \; 0
\end{equation}
\begin{equation}
    1 \wedge 1 \; := \; 1
\end{equation}

\smallskip

\begin{definition}
\end{definition}
\emph{disjunction:}

$ \vee : \Sigma \times \Sigma \mapsto \Sigma$:
\begin{equation}
    0 \vee 0 \; := \; 0
\end{equation}
\begin{equation}
    0 \vee 1 \; := \; 1
\end{equation}
\begin{equation}
    1 \vee 0 \; := \; 1
\end{equation}
\begin{equation}
    1 \vee 1 \; := \; 1
\end{equation}

\smallskip

\begin{definition}
\end{definition}
\emph{implication:}

$ \rightarrow : \Sigma \times \Sigma \mapsto \Sigma$:
\begin{equation}
    0 \rightarrow 0 \; := \; 1
\end{equation}
\begin{equation}
    0 \rightarrow 1 \; := \; 1
\end{equation}
\begin{equation}
    1 \rightarrow 0 \; := \; 0
\end{equation}
\begin{equation}
    1 \rightarrow 1 \; := \; 1
\end{equation}

\smallskip

\begin{definition}
\end{definition}
\emph{biimplication:}

$ \leftrightarrow : \Sigma \times \Sigma \mapsto \Sigma$:
\begin{equation}
    0 \leftrightarrow 0 \; := \; 1
\end{equation}
\begin{equation}
    0 \leftrightarrow 1 \; := \; 0
\end{equation}
\begin{equation}
    1 \leftrightarrow 0 \; := \; 0
\end{equation}
\begin{equation}
    1 \leftrightarrow 1 \; := \; 1
\end{equation}

\smallskip

In order of avoiding too many parentheses we will assume the
following (traditional) hierarchy  of strength for the logical
connectives:
\begin{enumerate}
    \item $ \neg $
    \item $ \wedge $ and $ \vee $
    \item $ \rightarrow $ and $ \leftrightarrow $
\end{enumerate}

\smallskip

Clearly:

\begin{theorem}
\end{theorem}
\begin{equation}
    ( x_{1} \leftrightarrow x_{2} ) \; \leftrightarrow  ( (x_{1}
    \rightarrow x_{2}) \; \wedge \; ( x_{2} \rightarrow x_{1}) )
\end{equation}
 Given a number $ n \in \mathbb{N}_{+} $ and a map $ f :
\Sigma^{n} \mapsto \Sigma $:
\begin{definition}
\end{definition}
\emph{f is a tautology:}
\begin{equation}
 f( x_{1} , \cdots , x_{n} ) \; = \; 1 \; \; \forall x_{1}, \cdots,
 x_{n} \in \Sigma
\end{equation}

We will adopt the following:
\begin{definition} \label{def:logical notation}
\end{definition}
\emph{logical notation:}
\begin{equation}
     f( x_{1} , \cdots , x_{n} )  \; := \; \text{ f is a tautology}
\end{equation}

It can be easily verified that:

\begin{theorem}
\end{theorem}
\emph{Double negation adfirms:}
\begin{equation}
   \neg \neg x \; \leftrightarrow \;  x
\end{equation}

\begin{theorem} \label{th:Principle of noncontradiction}
\end{theorem}
\emph{Principle of Noncontradiction:}
\begin{equation}
    \neg ( x \wedge \neg x)
\end{equation}

\smallskip

\begin{theorem} \label{th:De Morgan's laws}
\end{theorem}
\emph{De Morgan Laws:}
\begin{equation}
    x_{1} \wedge x_{2} \; \leftrightarrow \neg ( \neg x_{1} \vee \neg
    x_{2} )
\end{equation}
\begin{equation}
    x_{1} \vee x_{2} \; \leftrightarrow \; \neg ( \neg x_{1} \wedge \neg
    x_{2} )
\end{equation}

\smallskip

\begin{corollary}
\end{corollary}
\emph{Tertium non datur:}
\begin{equation}
    x \vee \neg x
\end{equation}
\begin{proof}
 The thesis follows applying the theorem \ref{th:De Morgan's laws}
 and the theorem \ref{th:Principle of noncontradiction}
\end{proof}

\smallskip

\begin{theorem} \label{th:Scotus theorem}
\end{theorem}
\emph{Scotus Theorem (ex absurdo quodlibet sequitur):}
\begin{equation}
    x_{1} \wedge \neg x_{1} \; \rightarrow \; x_{2}
\end{equation}
\begin{proof}
\begin{equation} \label{eq:auxialiary formula number 1}
    x_{1} \wedge \neg x_{1} \; \rightarrow \; x_{1}
\end{equation}
\begin{equation} \label{eq:auxialiary formula number 2}
    x_{1}  \wedge \neg x_{1} \; \rightarrow \; \neg x_{1}
\end{equation}
\begin{equation} \label{eq:auxialiary formula number 3}
  \neg x_{1}  \; \rightarrow \;  \neg x_{1} \vee x_{2}
\end{equation}
\begin{equation} \label{eq:auxialiary formula number 4}
   x_{1} \wedge  ( \neg x_{1} \vee x_{2} ) \; \rightarrow \; x_{2}
\end{equation}
Combining equation \ref{eq:auxialiary formula number 1}, equation
\ref{eq:auxialiary formula number 2}, equation \ref{eq:auxialiary
formula number 3} and equation \ref{eq:auxialiary formula number
4} the thesis follows
\end{proof}

It may be also easily verified that:

\begin{theorem} \label{th:reflectivity of implication}
\end{theorem}
\emph{Reflectivity of implication:}
\begin{equation}
    x \; \rightarrow \; x
\end{equation}

\smallskip

\begin{theorem} \label{th:transitivity of implication}
\end{theorem}
\emph{transitivity of implication:}
\begin{equation}
    (( x_{1} \rightarrow x_{2} ) \wedge  ( x_{2} \rightarrow x_{3}
    )) \; \rightarrow \;  ( x_{1} \rightarrow x_{3} )
\end{equation}

from which it follows that:

\begin{corollary}
\end{corollary}
\emph{Modus ponens:}
\begin{equation}
    x_{1} \wedge ( x_{1} \rightarrow x_{2} ) \; \rightarrow \; x_{2}
\end{equation}

\smallskip

Furthermore it may be easily verified that:
\begin{theorem} \label{th:Principle of Contraposition}
\end{theorem}
\emph{Principle of Contraposition:}
\begin{equation}
    ( x_{1} \rightarrow x_{2} ) \; \leftrightarrow \; ( \neg x_{2}
    \rightarrow \neg x_{1} )
\end{equation}

from which it follows that:

\begin{corollary}
\end{corollary}
\emph{Modus tollens:}
\begin{equation}
   \neg x_{2} \wedge ( x_{1} \rightarrow x_{2} ) \; \rightarrow \;
   \neg x_{1}
\end{equation}

\smallskip

Furthermore:
\begin{theorem}
\end{theorem}
\emph{Reductio ad absurdum:}
\begin{equation}
    (\neg x_{1} \rightarrow  x_{2} \wedge \neg x_{2} ) \;
    \rightarrow \; x_{1}
\end{equation}
\begin{proof}
  The thesis immediately follows combining theorem \ref{th:Principle of noncontradiction} and
  theorem \ref{th:Principle of Contraposition}
\end{proof}

\bigskip

Classical Propositional Calculus may be formalized as a formal
systems in many ways, for instance choosing as axiom the theorem
\ref{th:Principle of noncontradiction} and as rules of inference
the theorem \ref{th:reflectivity of implication}, theorem
\ref{th:transitivity of implication} and the theorem
\ref{th:Principle of Contraposition}
\newpage
\section{Notation}
\begin{center}
\begin{tabular}{|c|c|}
  \hline
   $ i.e. $ & id est \\
  $ \forall $ & for all (universal quantificator) \\
  $ \exists $ & exists (existential quantificator) \\
   $ G \, , \, \forall_{+} $ & future universal quantificator \\
   $ H \, , \, \forall_{-} $ & past universal quantificator \\
   $ F \, , \, \exists_{+} $ & future existential quantificator \\
   $ P \, , \,  \exists_{-} $ & past  existential quantificator \\
   $ x \; = \; y $ & x is equal to y \\
  $ x \; := \; y $ & x is defined as y \\
  $ \partial S $ & boundary of S \\
  $ I^{+} (S) $ & chronological future of the set S \\
  $ I^{-} (S) $ & chronological past of the set S \\
$ J^{+} (S) $ & causal future of the set S \\
$ J^{-} (S) $ & causal past of the set S \\
$ \mathcal{I}^{+} $ & future null infinity \\
$ \mathcal{I}^{-} $ & past null infinity \\
$ V_{chronology} ( M , g_{ab} )$ & chronology violating set of the spacetime $ ( M , g_{ab} ) $ \\
$ H_{chronology} ( M , g_{ab} )$ & chronology horizon of the spacetime $ ( M , g_{ab} ) $ \\
  $ \Sigma $ & binary alphabet \\
  $ \wedge $ &  conjunction \\
  $ \vee $ & disjunction \\
  $ \neg $ & negation \\
  $ \rightarrow $ & implication \\
  $ \leftrightarrow $ & biimplication \\
  \hline
\end{tabular}
\end{center}


\newpage
\begin{thebibliography}{10}

\bibitem{Hawking-Ellis-73}
S.W. Hawking~G.F.R. Ellis.
\newblock {\em The large scale structure of space-time}.
\newblock Cambridge University Press, Cambridge, 1973.

\bibitem{Wald-84}
R.M. Wald.
\newblock {\em General Relativity}.
\newblock The University of Chicago Press, Chicago, 1984.

\bibitem{Visser-96}
M.~Visser.
\newblock {\em Lorentzian Wormholes. From Einstein to Hawking}.
\newblock Springer, New York, 1996.

\bibitem{Visser-03}
M.~Visser.
\newblock The quantum physics of chronology protection.
\newblock In G.W. Gibbons E.P.S. Shellard~S.J. Rankin, editor, {\em The Future
  of Theoretical Physics and Cosmology}, pages 161--173. Cambridge University
  Press, Cambridge, 2003.

\bibitem{Hawking-92}
S.~Hawking.
\newblock The chronology protection conjecture.
\newblock {\em Phys. Rev. D}, 46:603--611, 1992.

\bibitem{Weinberg-95}
S.~Weinberg.
\newblock {\em The Quantum Theory of Fields, vol.1}.
\newblock Cambridge University Press, Cambridge, 1996.

\bibitem{Headrick-Gott-94}
M.P. Headrick~J.R. Gott.
\newblock (2+1)-{D}imensional {S}pacetimes {C}ontaining {C}losed {T}imelike
  {C}urves.
\newblock {\em Physical Review D}, 50:7244--7259, 1994.

\bibitem{Kay-Radzikowski-Wald-97}
B.S. Kay M.J. Radzikowki~R.M. Wald.
\newblock Quantum field theory on spacetimes with a compactly generated
  {C}auchy horizon.
\newblock {\em Commun. Math. Phys.}, 183:533--556, 1997.
\newblock gr-qc/9603012.

\bibitem{Kay-06}
B.S. Kay.
\newblock {Q}uantum {F}ield {T}heory in {C}urved {S}pacetime.
\newblock In J.P. Francoise G.L. Naber~T.S. Tsun, editor, {\em Encyclopedia of
  Mathematical Physics. Vol.4}, pages 202--216. Birkhauser, Basel, 2006.

\bibitem{Wald-94}
R.M. Wald.
\newblock {\em Quantum Field Theory in Curved Spacetime and Black Hole
  Thermodynamics}.
\newblock The University of Chicago Press, Chicago, 1994.

\bibitem{Deligne-Etingof-Freed-Jeffrey-Kazhdan-Morgan-Morrison-Witten-99a}
P.~Deligne P. Etingof D.S. Freed L.C. Jeffrey D. Kazhdan J.W.
Morgan D.R.
  Morrison~E. Witten.
\newblock {\em Quantum Fields and Strings: A Course for Mathematicians, vol.
  1}.
\newblock American Mathematical Society, 1999.

\bibitem{Deligne-Etingof-Freed-Jeffrey-Kazhdan-Morgan-Morrison-Witten-99b}
P.~Deligne P. Etingof D.S. Freed L.C. Jeffrey D. Kazhdan J.W.
Morgan D.R.
  Morrison~E. Witten.
\newblock {\em Quantum Fields and Strings: A Course for Mathematicians, vol.
  2}.
\newblock American Mathematical Society, 1999.

\bibitem{Rovelli-04}
C.~Rovelli.
\newblock {\em Quantum Gravity}.
\newblock Cambridge University Press, Candbridge, 2004.

\bibitem{Connes-98}
A.~Connes.
\newblock Noncommutative {G}eometry: {T}he {S}pectral {A}spect.
\newblock In A.~Connes K. Gawedzki~J. Zinn-Justin, editor, {\em Quantum
  Symmetries}, pages 643--686. Elsevier Science, Amsterdam, 1998.

\bibitem{Landi-97}
G.~Landi.
\newblock {\em An Introduction to Noncommutative Spaces and Their Geometries}.
\newblock Springer Verlag, Berlin, 1997.

\bibitem{Ambiorn-Carfora-Marzuoli-97}
J.~Ambiorn M. Carfora~A. Marzuoli.
\newblock {\em The Geometry of Dynamical Triangulations}.
\newblock Springer, Berlin, 1997.

\bibitem{Prugovecki-92}
E.~Prugove\v{c}ki.
\newblock {\em Quantum Geometry. A Framework for Quantum General Relativity}.
\newblock Kluwer Academic Publisher, Dordrecht, 1992.

\bibitem{Finkelstein-97}
D.R. Finkelstein.
\newblock {\em Quantum Relativity}.
\newblock Springer-Verlag, Berlin, 1997.

\bibitem{Prior-62}
A.N. Prior.
\newblock Tense-{L}ogic and the {C}ontinuity of {T}ime.
\newblock {\em Studia Logica}, 13, 1962.

\bibitem{Segre-06}
G.~Segre.
\newblock The multihistory approach to the time-travel paradoxes of {G}eneral
  {R}elativity: mathematical analysis of a toy model.
\newblock {\em math-ph/0610085}.

\bibitem{Mendelson-64}
E.~Mendelson.
\newblock {\em Introduction to Mathematical Logic}.
\newblock D. Van Nostrand Company, Princeton, 1964.

\end{thebibliography}
\end{document}